\documentclass[
reprint,           
superscriptaddress,
amsmath,           
amssymb,           
aps,               
prd,               
notitlepage,       
floatfix,          
nofootinbib,
]{revtex4-1}

\usepackage{tensor}     
\usepackage{float}
\usepackage[caption = false]{subfig} 
\usepackage[final]{graphicx}   
\usepackage[
colorlinks=true,        
citecolor=blue,         
linkcolor=blue,         
urlcolor=blue           
]{hyperref}             
\usepackage{bm}         
\usepackage{xcolor}     
\usepackage{lipsum}
\usepackage{color}      
\usepackage[utf8]{inputenc} 
\usepackage[section]{placeins} 
\usepackage{hyperref}

\newcommand{\nc}{\newcommand*}

\nc{\al}{\alpha}
\nc{\s}{\sigma}
\nc{\dt}{\delta}
\nc{\Dt}{\Delta}
\nc{\Ld}{\Lambda}
\nc{\p}{\partial}
\nc{\om}{\omega}
\nc{\Om}{\Omega}
\nc{\rd}{\mathrm{d}}
\nc{\Od}[1]{\mathcal{O}(#1)} 
\nc{\kp}{\kappa}

\def\({\left(}
\def\){\right)}
\def\[{\left[}
\def\]{\right]}
\def\e{\begin{equation}}
\def\q{\end{equation}}
\def\m{\begin{eqnarray}}
\def\n{\end{eqnarray}}
\nc{\Eq}[1]{Eq.~\eqref{#1}}     
\nc{\Fig}[1]{Fig.~\ref{#1}}     
\nc{\Table}[1]{Table~\ref{#1}}  
\nc{\Sec}[1]{Sec.~\ref{#1}}     
\nc{\Msun}{M_\odot}             
\nc{\fpbh}{f_{\mathrm{pbh}}}    
\nc{\fpbhn}{f_{\mathrm{pbh0}}}    
\nc{\mR}{\mathcal{R}} 
\nc{\seq}{\sigma_{\mathrm{eq}}}
\nc{\ogw}{\Omega_{\mathrm{GW}}}
\nc{\gpcyr}{\mathrm{Gpc}^{-3}\,\mathrm{yr}^{-1}}
\nc{\lvc}{LIGO/Virgo} 
\nc{\SNR}{\mathrm{SNR}} 
\nc{\mmin}{{m_{\mathrm{min}}}}
\nc{\mmax}{{m_{\mathrm{max}}}}
\nc{\Mmin}{{M_{\mathrm{min}}}}
\nc{\fmin}{{f_{\mathrm{min}}}}
\nc{\VT}{\mathrm{VT}}
\nc{\rhoGW}{\rho_{\mathrm{GW}}}
\nc{\vth}{\vec{\theta}}
\nc{\vd}{\vec{d}}
\nc{\vla}{\vec{\lambda}}
\nc{\Nobs}{N_{\mathrm{obs}}}
\nc{\av}[1]{\langle #1 \rangle} 
\nc{\km}{\mathrm{km}}
\nc{\Mpc}{\mathrm{Mpc}}
\nc{\Tobs}{T_{\mathrm{obs}}}
\nc{\Ntemp}{N_{\mathrm{temp}}}
\nc{\addref}{[\textcolor{red}{add ref}] } 
\nc{\eg}{\textit{e.g.~}}
\nc{\app}{\approx}
\nc{\hf}{\frac{1}{2}}
\nc{\discuss}{\textcolor{red}{Add discussion here!}}
\nc{\red}[1]{\textcolor{red}{#1}}
\nc{\mH}{\mathcal{H}}
\nc{\cs}{c_s^2}
\nc{\Sij}[1]{S_{ij}^{(#1)}}
\nc{\vi}[1]{v_i^{(#1)}}
\nc{\no}{\nonumber}
\def\<{\left\langle}
\def\>{\right\rangle}

\nc{\bk}{\bm{k}}
\nc{\bq}{\bm{q}}
\nc{\bp}{\bm{p}}
\nc{\bl}{\bm{l}}
\nc{\bx}{\bm{x}}
\nc{\be}{\mathbf{e}}
\nc{\mS}{\mathcal{S}}
\nc{\te}{\tilde{\eta}}
\nc{\tp}{\tilde{p}}
\nc{\tk}{\tilde{k}}
\nc{\tx}{\tilde{x}}
\nc{\tF}{\tilde{F}}
\nc{\tA}{\tilde{A}}
\nc{\mkpq}{|\bk-\bp-\bq|}
\nc{\mpq}{|\bp-\bq|}
\nc{\mkp}{|\bk-\bp|}
\nc{\mSi}[1]{\mS^{(#1)}({\bk, \eta})}
\nc{\vk}{\vec{k}}
\nc{\kstar}{k_*}
\nc{\fstar}{f_*}
\nc{\xstar}{x_*}
\nc{\mpbh}{m_{\rm{pbh}}}
\nc{\bn}[1]{\bm{n}_{\text{#1}}}
\nc{\bC}[1]{\bm{C}_{\text{#1}}}
\nc{\NTOA}{N_{\text{TOA}}}
\nc{\Nmode}{{N_{\text{mode}}}}
\nc{\ARN}{A_{\rm{RN}}}
\nc{\gRN}{\gamma_{\rm{RN}}}
\nc{\bS}{\mathbf{\Sigma}}
\nc{\br}{\mathbf{r}}
\nc{\bN}{\mathbf{N}}

\nc{\arXiv}[2]{\href{http://arxiv.org/pdf/#1}{{\tt [#2/#1]}}}
\nc{\arXivold}[1]{\href{http://arxiv.org/pdf/#1}{{\tt [#1]}}}

\renewcommand{\vec}[1]{\boldsymbol{#1}} 

\begin{document}
	
\title{Near-horizon microstructure and superradiant instability of black holes}
	
\author{Rong-Zhen Guo}
\email{guorongzhen@itp.ac.cn}
\affiliation{CAS Key Laboratory of Theoretical Physics,
	Institute of Theoretical Physics, Chinese Academy of Sciences,
	Beijing 100190, China}
\affiliation{School of Physical Sciences,
	University of Chinese Academy of Sciences,
	No. 19A Yuquan Road, Beijing 100049, China}
\author{Chen Yuan}
\email{yuanchen@itp.ac.cn}
\affiliation{CAS Key Laboratory of Theoretical Physics,
Institute of Theoretical Physics, Chinese Academy of Sciences,
Beijing 100190, China}
\affiliation{School of Physical Sciences,
University of Chinese Academy of Sciences,
No. 19A Yuquan Road, Beijing 100049, China}
\author{Qing-Guo Huang}
\email{corresponding author: huangqg@itp.ac.cn}
\affiliation{CAS Key Laboratory of Theoretical Physics,
	Institute of Theoretical Physics, Chinese Academy of Sciences,
	Beijing 100190, China}
\affiliation{School of Physical Sciences,
	University of Chinese Academy of Sciences,
	No. 19A Yuquan Road, Beijing 100049, China}
\affiliation{School of Fundamental Physics and Mathematical Sciences
Hangzhou Institute for Advanced Study, UCAS, Hangzhou 310024, China}
\affiliation{Center for Gravitation and Cosmology,
	College of Physical Science and Technology,
	Yangzhou University, Yangzhou 225009, China}
	
\date{\today}

\begin{abstract}

Ultralight bosons, as important candidates of dark matter, can condense around spinning black holes (BHs) to form long-lived ``boson clouds'' due to superradiance instability. The boson-BH system can be observed through gravitational wave detection and may become a new window to find traces of ultralight bosons. In this letter we explore the effects on the superradiant instability of BHs from the near-horizon microstructure. By introducing the reflection parameter near a BH horizon, we derived analytical results on the corrections to both energy levels of bosonic cloud and its characteristic frequencies of superradiance instability. Our results imply that the evolution of a boson-BH system and gravitational waves it emits would be influenced by the near-horizon physics of a BH.
\end{abstract}
	
\pacs{???}
	
\maketitle


{\it Introduction. }
Ultralight bosonic particles, which are predicted in various beyond Standard Model scenarios \cite{Arvanitaki:2009fg,Essig:2013lka,Irastorza:2018dyq,Goodsell:2009xc,Jaeckel:2010ni,Graham:2015rva,Agrawal:2018vin}, could be a promising dark matter (DM) candidate \cite{Preskill:1982cy,Abbott:1982af,Dine:1982ah,svrcek2006axions,Arvanitaki:2009fg,Arvanitaki:2010sy,Essig:2013lka,Brito:2015oca,Marsh:2015xka,Hui:2016ltb,Annulli:2020lyc,Chadha-Day:2021szb}.
Given that detecting ultralight particles is challenging in particle physics, using gravitational waves (GWs) to probe ultralight bosons has aroused much attention recently. The main mechanism responsible for this method is the superradiant instability in a massive boson and spinning black hole (BH) system \cite{Damour:1976kh,Zouros:1979iw,Detweiler:1980uk,Dolan,string_axiverse,Shlapentokh-Rothman:2013ysa,Pani:2012vp,Pani:2012bp,Witek:2012tr,sr_tensor,Endlich:2016jgc,East:2017mrj,East:2017ovw,sr_vector_4,Cardoso:2018tly,East:2018glu,Frolov:2018ezx,Dolan:2018dqv,Baumann:2019eav,Brito:2020lup,Tsukada:2020lgt,Yuan:2021ebu}. When the Compton wavelength of bosons is about the order of the BH horizon, the bosons would extract energy and angular momentum from the BH, forming a macroscopic rotating boson cloud, which dissipates its energy through the emission of nearly monochromatic GWs on a very lone timescale. Therefore, the boson-BH system can act as a continuous GW source and could be detected by GW observations \cite{Arvanitaki:2014wva,Arvanitaki:2016qwi,Baryakhtar:2017ngi,Brito:2017wnc,Brito:2017zvb,Isi:2018pzk,Ghosh:2018gaw,Palomba:2019vxe,Sun:2019mqb,Zhu:2020tht,Brito:2020lup,Ng:2020jqd,Tsukada:2020lgt,Yuan:2021ebu}.
Since ultralight bosons such as axions or string axiverse could be important candidates of dark
matter  and it is difficult
to find its evidence by traditional collision experiments, superradiant
instabilities of astrophysical BHs have been used to constrain its
parameters \cite{Kodama:2011zc,Brito:2014wla,Arvanitaki:2010sy,Brito:2017wnc,Brito:2017zvb,Tsukada:2018mbp,Zhu:2020tht,Tsukada:2020lgt}.

BH in general relativity could be seen as a special
region in the spacetime surrounded by event horizon, which behaves
as a smooth one-way membrane \cite{Thorne:1986iy}. Because of Penrose-Hawking
singularity theorems \cite{Hawking:1973uf}, there exists a physically
ill-defined singularity inside its event horizon. However, this picture
is challanged by two aspects. One aspect is the existence of singularity
for the belief that there should be no such pathological point in
the real physical world. To avoid the formation of singularity, the
idea that BH is not the end state of gravitational collapse arises
\cite{Barcelo:2007yk}. For example, the end state might be a gravastar
\cite{Morris:1988tu}, boson star \cite{Schunck:2003kk}, wormhole
\cite{Morris:1988tu} or other exotic compact object (ECO) \cite{Holdom:2016nek}.
Another aspect stems from the comprehensive consideration of both
gravity and quantum effects. Several inferences of BH thermodynamics
like Bekenstein-Hawking entropy imply the discrete degrees of freedom
living on event horizon, which lead to the BH area quantization (named
Bekenstein-Mukhanov anstaz) \cite{Bekenstein:1974jk,Bekenstein:1995ju}.
Furthermore, several candidates of quantum gravity predict different
quantum-corrected BH models (e.g. fuzzball \cite{skenderis2008fuzzball}
in string theory). Moreover, for solving the BH information paradox
\cite{polchinski2017black}, firewall \cite{almheiri2013black} might
exist near event horizon. All of the above would significantly change
the physics near the event horizon.

Despite the existence of colorful quantum-corrected BH models or ECOs,
these modifications to classical BHs can be described by a phenomenological
model that introduces reflectivity $\mathcal{R}(\omega)$ near the
horizon of a BH. It is widely used in the study of echoes of gravitational
waves \cite{Cardoso:2016rao,Cardoso:2019apo,mark2017recipe,Price:2017cjr,Fang:2021iyf}.
Cardoso et al. \cite{Cardoso:2019apo} has proved BH echoes from BH
area quatization. It has been also used to test possible Lorentz violation
near event horizon \cite{Oshita:2018fqu}.
This effective description could be understood from two different
perspectives. Mathematically, such a boundary condition is essentially
the linearization of much general boundary conditions \cite{Burgess:2018pmm}.
And from the physical point of view, because the radial equation is
a Schrodinger-like equation, the reflection parameters $\mathcal{R}(\omega)$
here can be considered to mark the nonzero probability of particle
reflection near the black hole horizon. In addition, Burgess et.al
\cite{Rummel:2019ads,Burgess:2018pmm} point that $\mathcal{R}(\omega)$
is related to the couplings of effective field theory living in the
near-horizon region, thus it could be used to constrain fundamental physics
in a more direct way.

Since superradiance instability could be solved as a boundary value
problem in the framework of BH perturbation theory \cite{brito2015superradiance},
it implies that superradiance instability could be used to detect
possible effects of the near horizon microstructure, thus provide
clues of quantum gravity. Here we study a scalar field minically coupled
to gravity on the Kerr background with the modified boundary condition
given near event horizon and derive analytical results of the corrections to the bound state frequencies of bosonic clouds. In this letter we take the units $c=G=1$.



{\it Superradiant instability of black holes. } We consider a scalar field $\psi$ with mass $\mu$ in the vicinity of a Kerr black hole whose metric in standard Boyer-Lindquist coordinates $\left\{ t,r,\theta,\phi\right\}$ takes the following form
\m
ds^{2}=-\left(1-\frac{2Mr}{\Sigma}\right)dt^{2}-\frac{2Mar\sin^{2}\theta}{\Sigma}dtd\phi \nonumber \\
+\frac{\Sigma}{\Delta}dr^{2}+\Sigma d\theta^{2}+\frac{A\sin^{2}\theta}{\Sigma}d\phi^{2},
\n
where $M$ and $J=aM$ are the mass and angular momentum of Kerr black hole, and
\m
\Sigma&=&r^{2}+a^{2}\cos^{2}\theta, \\
\Delta&=&r^{2}-2Mr+a^{2}\equiv\left(r-r_{+}\right)\left(r-r_{-}\right), \\
A&=&(r^{2}+a^{2})-a^{2}\Delta\sin^{2}\theta.
\n
The event horizon and Cauchy horizon are located at $r_+=M+\sqrt{M^2-a^2}$ and $r_-=M-\sqrt{M^2-a^2}$, respectively.
The Klein-Gordon field equation $(\nabla^{\nu}\nabla_{\nu}-\mu^2) \psi=0$ in such a background reads
\begin{equation}
\begin{split} & \left[\frac{\left(r^{2}+a^{2}\right)^{2}}{\Delta}-a^{2}\sin^{2}\theta\right]\frac{\partial^{2}\psi}{\partial t^{2}}+\frac{4Mar}{\Delta}\frac{\partial^{2}\psi}{\partial t\partial\phi}\\
 & +\left[\frac{a^{2}}{\Delta}-\frac{1}{\sin^{2}\theta}\right]\frac{\partial^{2}\psi}{\partial\phi^{2}}-\frac{\partial}{\partial r}\left(\Delta\frac{\partial\psi}{\partial r}\right)\\
 & -\frac{1}{\sin\theta}\frac{\partial}{\partial\theta}\left(\sin\theta\frac{\partial\psi}{\partial\theta}\right)-\text{\ensuremath{\mu^{2}\Sigma^{2}\psi=0}}
\end{split}
\label{kle}
\end{equation}
whose mode solution, \cite{shlapentokh2015quantitative}, takes the form
\begin{equation}
\psi(t,r,\theta,\phi)=e^{-i\omega t}e^{im\phi} Z_{lm}(r) S_{lm}(\theta)
\end{equation}
with $\omega\in\mathbb{C}$ and azimuthal number $m\in\mathbb{Z}$, and then Eq.~(\ref{kle}) can be separated into
\m
\frac{1}{\sin\theta}\frac{d}{d\theta}\left(\sin\theta\frac{dS_{lm}}{d\theta}\right)+a^{2}\left(\omega^{2}-\mu^{2}\right)\cos^{2}\theta S_{lm} \nonumber\\
-\left(\frac{m^{2}}{\sin^{2}\theta}-\lambda\right)S_{lm}=0,
\label{slm}
\n
and
\m
\frac{d}{dr}\left(\Delta\frac{dZ_{lm}}{dr}\right)&+& \frac{\omega^{2}\left(r^{2}+a^{2}\right)^{2}-4aMm\omega r+a^{2}m^{2}}{\Delta} Z_{lm}  \nonumber \\
 &-&\left(\mu^{2}r^{2}+a^{2}\omega^{2}+\lambda\right) Z_{lm}=0,
 \label{zlm}
\n
where $\lambda$ is the separation constant. The radial equation, Eq.~(\ref{zlm}), is a confluent Heun equation, and we could rewrite it into the form for convenience \cite{Baumann2019}:
\m
\begin{aligned}
0=&\frac{1}{Z_{lm} \Delta} \frac{\mathrm{d}}{\mathrm{d} r}\left(\Delta \frac{\mathrm{d} Z_{lm}}{\mathrm{~d} r}\right) -\frac{\lambda}{\Delta}-\left(\mu^2-\omega^{2}\right)+\frac{P_{+}^{2}}{\left(r-r_{+}\right)^{2}}\\
&+\frac{P_{-}^{2}}{\left(r-r_{-}\right)^{2}}
-\frac{A_{+}}{\left(r_{+}-r_{-}\right)\left(r-r_{+}\right)}+\frac{A_{-}}{\left(r_{+}-r_{-}\right)\left(r-r_{-}\right)},
\label{confluent}
\end{aligned}
\n
where we have defined the following coefficients
\m
A_{\pm} \equiv P_{+}^{2}+P_{-}^{2}+\gamma^{2}+\gamma_{\pm}^{2} , ~ P_{\pm} \equiv \frac{a m-2M \omega r_{\pm}}{r_{+}-r_{-}}
\n
with
\m
\begin{aligned}
\gamma^{2} & \equiv \frac{1}{4}\left(r_{+}-r_{-}\right)^{2}\left(\mu^2-\omega^{2}\right) \\
\gamma_{\pm}^{2} & \equiv\left[M^{2}\left(\mu^2-7 \omega^{2}\right) \pm M\left(r_{+}-r_{-}\right)\left(\mu^2-2 \omega^{2}\right)\right].
\end{aligned}
\n

The eigenfunctions $S(\theta)$ in Eq.~(\ref{slm}) are the spheroidal harmonics with its eigenvalues \cite{simone1992massive}
\begin{equation}
\lambda\equiv\lambda^{lm}={\displaystyle \sum_{j}}f_{2j}^{lm}(ad)^{2j},\ \{j\in\mathbb{\mathbb{N}}\}
\label{lambda}
\end{equation}
where $(l,m)\in\mathbb{Z}$, $|m|\leq l$, $d^{2}=\omega^{2}-\mu^{2}$ , and $\lambda\simeq f_{0}^{lm}=l(l+1)$ at the leading order.

In this letter we mainly focus on the ultralight bosons with $\omega\sim \mu \ll 1/M$. Note that $\lambda$ itself is a complex number, and then even if we choose $\lambda=l(l+1)$, $l$ should be seen as a complex number close to a integer. We use the matched asymptotic expansion to get the spectrum of bosonic field analytically \cite{Starobinsky:1973aij,Baumann2019}.
Radial Teukolsky equation could be solved in ``far zone'' and ``near zone'' on the Kerr background. In the ``far zone'', which is dominated by the non-relativistic effect, the Klein-Gorden equation could be approximated to the Schrodinger equation with a Newtonian potential, so that the radial equation will have the same form of the radial hydrogen atom wave function:
\begin{equation}
\frac{d^{2}}{dr^{2}}\left(rZ_{lm}\right)+\left[\omega^{2}-\mu^{2}+\frac{2M\mu^{2}}{r}-\frac{l(l+1)}{r^{2}}\right](rZ_{lm})=0.\label{eq:15}
\end{equation}
Obviously, the far zone should be defined on $r\gg M$. For the pure outgoing wave condition at $kr\rightarrow +\infty$, the solution of the above equation reads
\begin{equation}
rZ_{lm}(r)=(2kr)^{l+1}e^{-kr}U(l+1-M\mu^{2}/k,2l+2,2kr)
\label{sggm}
\end{equation}
where $U$ is the confluent hypergeometric function, and $k=\sqrt{\mu^2-\omega^2}$.
Eq.~(\ref{sggm}) is just the bound state of hydrogen atoms if $l+1-M\mu^{2}/k\equiv l+1-\nu=-n$ ($n\in\text{\ensuremath{\mathbb{N}}}$). Here the state with superradiance instability could be approximately described by a small shift of energy level of hydrogen atoms, namely
\begin{equation}
\nu-l-1=n+\delta\nu,\ n\in\mathbb{N},\ \delta\nu\in\mathbb{C},\ \left|\delta\nu\right|\ll n.
\end{equation}
If we take an asymptotic expansion of Eq.~(\ref{eq:15}) at $2kr\to0$,
using the property of conflunt hypergeometric functions and gamma
functions:

\begin{equation}
\begin{aligned} & U(x,y,\alpha)=\frac{\Gamma(1-y)}{\Gamma(x-y+1)}\sum_{k=0}^{\infty}\frac{(x)_{k}\alpha^{k}}{(y)_{k}k!}\\
 & +\frac{\Gamma(y-1)}{\Gamma(x)}\alpha^{1-y}\sum_{k=0}^{\infty}\frac{(x-y+1)_{k}\alpha^{k}}{(2-y)_{k}k!};\ y\notin\boldsymbol{Z}
\end{aligned}
\label{eq:20}
\end{equation}

\begin{equation}
\Gamma(\alpha)\propto\frac{(-1)^{n}(1+O(\alpha+n))}{n!(\alpha+n)};\ (\alpha\rightarrow-n)\wedge n\in\mathbb{N}
\end{equation}
where $(x)_{0}\equiv1,\ (x)_{k}\equiv\text{\ensuremath{\Gamma\left(a+k\right)/\Gamma\left(a\right)}},$
the asymptotic expansion of $Z_{lm}(r)$ in Eq.~(\ref{sggm}) at $M\ll r\ll 1/k$ takes the form of
\begin{equation}
\begin{aligned}Z_{lm}(r) & \sim(-1)^{n}\frac{(2l+1+n)!}{(2l+1)!}(2kr)^{l}+\cdots\\
 & +(-1)^{n+1}\delta\nu(2l)!n!(2kr)^{-l-1}+\cdots\ .
\end{aligned}
\label{zlmkr0}
\end{equation}
On the other hand, in the ``near zone'', which should close to the event horizon at $r_+$, the main term of Radial Teukolsky equation is $-l(l+1)/\Delta$ and $P_{+}^{2}/\left(r-r_{+}\right)^{2}$, which could be easily read by Eq.~(\ref{confluent}). Introduce the rescale radial coordinate:
\begin{align}
z & =\left(r-r_{+}\right)/\left(r_{+}-r_{-}\right),
\end{align}
then Eq.~(\ref{zlm}) can be written as
\m
&& z\left(z+1\right)\frac{d}{dz}\left[z(z+1)\frac{dZ_{lm}}{dz}\right] \nonumber \\
&&+\left[P_{+}^{2}-l\left(l+1\right)z\left(z+1\right)\right]Z_{lm}=0.
\label{zlmllmu}
\n
Meanwhile, once fixed the radial coordinate, expanding Eq.~(\ref{confluent}) to the leading order of $\mu M$ could lead to:
\m
\begin{aligned}
&\left[\frac{\mathrm{d}^{2}}{\mathrm{~d} z^{2}}+\left(\frac{1}{z}+\frac{1}{z+1}\right) \frac{\mathrm{d}}{\mathrm{d} z}-\frac{\ell(\ell+1)}{z(z+1)}\right]Z_{lm}\\
&+\left[\frac{P_{+}^{2}}{z^{2}}+\frac{P_{+}^{2}}{(z+1)^{2}}-\frac{2 P_{+}^{2}}{z}+\frac{2 P_{+}^{2}}{z+1}\right]Z_{lm}=0
\label{zlmapp}
\end{aligned}
\n
If we compare Eq.~(\ref{zlmllmu}) and Eq.~(\ref{zlmapp}), we could estimate that Eq.~(\ref{zlmllmu}) should be valid on $r_+<r \ll \max(l/\omega,l/m)$. So these two regions have the overlap on $M\ll r \ll \max(l/\omega,l/m)$.
In order to match the solution in Eq.~(\ref{zlmkr0}) in the overlapping region $M\ll r \ll \max(l/\omega,l/m)$, the solution of Eq.~(\ref{zlmllmu}) is given by
\begin{equation}
Z_{lm}(z)=\left(\frac{z}{z+1}\right)^{iP_+}\left[Aw_{5}\left(z\right)+Bw_{6}\left(z\right)\right]
\end{equation}
where
\begin{equation}
\begin{aligned} & w_{5}(z)=\left[-\left(z+1\right)\right]^{l}{}_{2}F_{1}(-l,-l+2iP_+;-2l;\left(z+1\right)^{-1})\\
 & \sim\left(-z\right)^{l}\sim\left(\frac{-r}{r_{+}-r_{-}}\right)^{l}
\end{aligned}
\end{equation}
and

\m w_{6}\left(z\right)&&=\left[-\left(z+1\right)\right]^{-l-1}\no\\
	&&\times{}_{2}F_{1}(l+1,l+1+2iP_+;2l+2;\left(z+1\right)^{-1})\no\\
 && \sim\left(-z\right)^{-l-1}\sim\left(\frac{-r}{r_{+}-r_{-}}\right)^{-l-1},
\n
are two linear independent solution of hypergeometric equation. So that we find the coefficients $A,B$
are:
\begin{equation}
A=\left(-1\right)^{n+l}\left[2k(r_{+}-r_{-})\right]^{l}\frac{\left(2l+n+1\right)!}{\left(2l+1\right)!}
\end{equation}

\begin{equation}
B=\left(-1\right)^{n+l}\left[2k(r_{+}-r_{-})\right]^{-l-1}\delta\text{\ensuremath{\nu}}\left(2l\right)!\left(n\right)!.
\end{equation}
This is the matching condition gluning two asymptotic regions.


In order to figure out the solution of Eq.~(\ref{zlm}), we still need to take the boundary condition near horizon into account. At classical level, we should only have an ingoing wave at the horizon because the effective potential drops to zero. Here we suppose that the microstructure due to the quantum effects modify the classical description of a black hole at horizon scales and such effects can be encoded in a frequency dependent reflectivity $\mathcal{R}(\omega)$:
\begin{equation}
Z_{lm}(r_{*})\sim e^{-i\left(\omega-m\Omega_{H}\right)r_{*}}+\mathcal{R}(\omega)e^{i\left(\omega-m\Omega_{H}\right)r_{*}},
\label{eq:boundary}
\end{equation}
where
\m
\Omega_H= a/(2Mr_+),
\n
for $r_{*}\rightarrow -\infty$, where $r_*$ is the tortoise coordinate
\e
r_*=r+{2Mr_+\over r_+-r_-} \ln {r-r_+\over 2M}-{2Mr_-\over r_+-r_-} \ln {r-r_-\over 2M}.
\q
Actually $w_5$ and $w_6$ can be re-written by another set of function bases of hypergeometric equation $w_3$ and $w_4$ for respectively referring to ingoing and outgoing solutions near horizon, namely
\m
w_5&=&T_{53} w_3+T_{54}w_4,\\
w_6&=&T_{63} w_3+T_{64}w_4,
\n
where
\m
w_{3}(z)&=&{}_{2}F_{1}(-l,l+1;2iP_++1;-z),\\
w_{4}(z)&=&{}_{2}F_{1}(l+1-2iP_+,-l-2iP_+;-2iP_++1;-z),\no\\
\n
and
\m
T_{53}&=&\frac{(-1)^{l}\Gamma(1-2iP_+)\Gamma(2iP_+)\Gamma(-2l)}{\text{\ensuremath{\Gamma(-l)\Gamma(-l-2iP_+)\Gamma(1+2iP_+)}}}, \\
T_{54}&=&-\frac{(-1)^{l}\Gamma(2iP_+)\Gamma(-2l)}{\Gamma(-l+2iP_+)\Gamma(-l)},\\
T_{63}&=&-\frac{(-1)^{l}\Gamma(1-2iP_+)\Gamma(2iP_+)\Gamma(2l+2)}{\text{\ensuremath{\Gamma(l+1)\Gamma(l+1-2iP_+)\Gamma(1+2iP_+)}}},\\
T_{64}&=&\frac{(-1)^{l}\Gamma(2iP_+)\Gamma(2l+2)}{\Gamma(l+1+2iP_+)\Gamma(l+1)}.
\n
And then the boundary condition in Eq.~(\ref{eq:boundary}) gives
\begin{equation}
\mathcal{R}=\frac{AT_{54}+BT_{64}}{AT_{53}+BT_{63}}.
\end{equation}
Supposing $\mathcal{R}=\mathcal{R}_{0}e^{i\phi_{\mathcal{R}}}$, after a tedious but straightforward computation, we obtain
\begin{equation}
\delta\nu
=\delta\bar{\nu} \frac{1-\mathcal{R}_{0}e^{i\left(\phi_{\mathcal{R}}+2\sum_{j=1}^{l}\phi_{j}\right)}}{1+\mathcal{R}_{0}e^{i\left(\phi_{\mathcal{R}}+2\sum_{j=1}^{l}\phi_{j}\right)}},
\label{eq:deltanu}
\end{equation}
where $\phi_{j}=\arctan (2P_+/j)$, and
\begin{equation}
\begin{alignedat}{1}\delta \bar{\nu}=iP_+\left[2k(r_{+}-r_{-})\right]^{2l+1}\frac{(2l+n+1)!}{n!}\\
\left[\frac{l!}{(2l)!(2l+1)!}\right]^{2}\prod_{j=1}^{l}(j^{2}+4P_+^{2})
\end{alignedat}
\end{equation}
denotes the result without reflection at horizon \cite{Pani:2012bp}.

Considering $M\mu^2/\sqrt{\mu^2-\omega^2}=\nu$ and $\nu=\nu_0+\delta\nu$ with $\nu_0=n+l+1$, we get
\e
\omega=\omega_0+\delta\omega,
\q
where
\m
\omega_0^2&=&\mu^2-\mu^2 \({M\mu\over l+n+1}\)^2,\\
\delta\omega&\simeq& \frac{\delta\nu}{M}\left(\frac{M\mu}{l+n+1}\right)^3.
\n
Since the real and imaginary parts of $\delta\omega$ denote the level shift of massive scalar field and the stability/instability property of the corresponding mode, we decompose $\delta\omega$ on the Cartesian complex plane as follows
\m\label{qnm}
\text{Re}[\delta\omega] &=& \frac{\delta\bar{\nu}}{iM}\left(\frac{M\mu}{l+n+1}\right)^{3} \frac{2\mathcal{R}_{0}\sin\phi_\omega}{\mathcal{R}_{0}^{2}+1+2\mathcal{R}_{0}\cos \phi_\omega},
\label{51} \\
\text{Im}[\delta\omega] &=& \frac{\delta\bar{\nu}}{iM}\left(\frac{M\mu}{l+n+1}\right)^{3} \frac{1-\mathcal{R}_{0}^{2}}{\mathcal{R}_{0}^{2}+1+2\mathcal{R}_{0}\cos\phi_\omega},
\label{50}
\n
where $\phi_\omega=\phi_{\mathcal{R}}+2\text{\ensuremath{\sum_{j=1}^{l}\phi_{j}}}$.


{\it  Discussion.}
In this letter, we derive analytical formulas of superradiance instability influenced by the modified boundary conditions near the event horizon. Our result shows that the behavior of superradiance instability of ultralight scalar bosons could be affected by the potential near-horizon new physics. Since the evolution of a boson-BH system is related to the bound state frequencies, Eq.~(\ref{qnm}), our results imply that the properties such as the e-folding time a boson cloud takes to reach the saturation point, the maximum boson cloud mass, the gravitational wave signal would be influenced by the near-horizon microstructure.
Further astrophysical observation could be used to detect both the existence of ultralight bosons and the possible breakdown of classical BHs.

{\it Acknowledgments. }
This work is supported by the National Key Research and Development Program of China Grant No.2020YFC2201502, grants from NSFC (grant No. 11975019, 11690021, 11991052, 12047503),  the Key Research Program of the Chinese Academy of Sciences (Grant NO. XDPB15), Key Research Program of Frontier Sciences, CAS, Grant NO. ZDBS-LY-7009, and the science research grants from the China Manned Space Project with NO. CMS-CSST-2021-B01.

\bibliography{reference}

\end{document}